\documentclass[12pt,a4paper]{article}
\usepackage{graphicx}
\usepackage{xcolor}
\usepackage{comment}
\usepackage{subcaption}
\usepackage{geometry}
\usepackage{tikz-cd}
\usepackage{amsmath}
\usepackage{amssymb}
\usepackage{physics}
\usepackage{cite}
\usepackage[english]{babel}
\usepackage[utf8]{inputenc}
\usepackage{bm}
\usepackage{braket}
\usepackage[colorlinks=true, allcolors=blue]{hyperref}
\usepackage{fancyhdr}
\usepackage{cleveref}

\usepackage{titlesec}
\titleformat{\section}
  {\fontsize{14}{12}\bf}{\thesection}{1em}{}
\titleformat{\subsection}
  {\fontsize{12}{10}\bf}{\thesubsection}{1em}{}
\newgeometry{top=3cm,bottom=2.5cm,left=2.5cm,right=2.5cm}
\renewcommand{\gcd}{\mathrm{gcd}}

\begin{document}
%\maketitle
\begin{titlepage}
\thispagestyle{empty}
\phantom{fez}

\vspace{-0.5cm}

\begin{center}

{\Huge {\bf $SL_2(\mathbb{R})$ symmetries of SymTFT and non-invertible $U(1)$ symmetries of Maxwell theory}}

\vspace{0.5cm}

{\Large Azeem Hasan, Shani Meynet, \\ and Daniele Migliorati}

\vspace{1cm}

{\footnotesize

\textit{Department of Mathematics, Uppsala University\\
Box 480, SE-75106 Uppsala, Sweden} \\

\textit{Centre for Geometry and Physics, Uppsala University \\
Box 480, SE-75106 Uppsala, Sweden} \\

}

\footnotesize{{\tt \href{mailto:azeem.hasan@math.uu.se}{azeem.hasan@math.uu.se}}}, \footnotesize{{\tt \href{mailto:shani.meynet@math.uu.se}{shani.meynet@math.uu.se}}}\\
\footnotesize{{\tt \href{mailto:daniele.migliorati@math.uu.se}{daniele.migliorati@math.uu.se}}}
\vspace{1cm}

{\large \textbf{Abstract}}

\end{center}

\noindent Recent proposals for the Symmetry Topological Filed Theory (SymTFT) of Maxwell theory admit a 0-form symmetry compatible with the classical $SL_2(\mathbb{R})$ duality of electromagnetism. We describe how to realize these automorphisms of the SymTFT in terms of its operators and we describe their effects on the dynamical theory and its global variants. In the process, we show that the classical $U(1)$ symmetry, corresponding to the the stabilizer of $SL_2(\mathbb{R})$, can be restored as a non-invertible one, by means of an infinite series of discrete gauging. This provide an example of the reemergence of a classical symmetry in the quantum regime, which was not broken by anomalies, but rather by the quantization of electromagnetic fluxes. However, this procedure comes at the price of introducing “continuous" condensates that trivialize all line operators.

\vfill

\end{titlepage}

\setcounter{page}{1}
{\hypersetup{linkcolor=black}
\tableofcontents
}

\section{Introduction}
Symmetries have been the most fruitful source of exact results in the study of quantum field theory. For most of this endeavor, they have also been a settled concept: symmetries are transformations that leave a physical system invariant. Such transformations necessarily form a group and, following Wigner, fields and quantum states are organized according to the representations of this group. However, in the past decade or so, this solid concept has become surprisingly fluid.

The newfound activity stems from the observation that symmetries can be realized by operators that depend topologically on their support \cite{Gaiotto:2014kfa}. The flurry of current development comes from turning this relationship on its head and defining symmetries in a theory to be its topological operators. Topological operators in a theory can be composed in more elaborate ways than standard group multiplication. The correct mathematical framework to describe these generalized symmetries is believed to be that of (higher) fusion categories (see \cite{Cordova:2022ruw,McGreevy:2022oyu,Schafer-Nameki:2023jdn,Brennan:2023mmt, Bhardwaj:2023kri ,Shao:2023gho, Carqueville:2023jhb} for a review and \cite{Bhardwaj:2017xup, Copetti:2023mcq, Bhardwaj:2023ayw, Bartsch:2023wvv, Bonetti:2024cvq} for recent developments).

An important conceptual underpinning of this paradigm is the symmetry topological field theory (SymTFT) \cite{Ji:2019jhk ,Gaiotto:2020iye, Apruzzi:2021nmk, Baume:2023kkf, Freed:2022qnc}. For a $d$-dimensional theory $\mathcal{T}$, the SymTFT is an auxiliary topological theory in $d+1$ dimensions that encodes the symmetries of $\mathcal{T}$ as well as their categorical structure. In this framework, the theory $\mathcal{T}$ can be realized by considering a bulk-boundary system. As bulk, we take the SymTFT on $X_{d+1} = M_{d} \times I$, with $I$ a closed interval, while at the two ends of the interval $I$ we take two different boundary theories coupled to the SymTFT. One of them is a relative theory $\hat{\mathcal{T}}$ and the other is a topological theory that fixed boundary condition $\mathcal{B}$ for the fields of the SymTFT. With a suitable choice of $\hat{\mathcal{T}}$ and $\mathcal{B}$, the theory $\mathcal{T}$ can be recovered by shrinking the interval $I$. Topological operators of the SymTFT correspond to symmetry operators of $\mathcal{T}$, and symmetries of the SymTFT correspond to various operations on the topological boundary, manipulations that connect $\mathcal{T}$ with other global variants of the same dynamical theory. In particular, to this second kind of operators belongs the gauging of finite symmetries of $\mathcal{T}$ or the stacking of some SPT for a symmetry of $\mathcal{T}$. In this sense the SymTFT decouples the topological aspects (i.e. symmetries) of $\mathcal{T}$ from the detailed dynamics \cite{Aharony:2013hda, Kapustin:2014gua,Hsin:2018vcg, Gaiotto:2020iye, Gukov:2020btk,Kong:2020jne,Ji:2021esj,DelZotto:2022ras, Kaidi:2022cpf, Bashmakov:2022uek, Kaidi:2023maf, Chen:2023qnv, Bashmakov:2023kwo, Sun:2023xxv, Cordova:2023bja, Bhardwaj:2023idu,Bhardwaj:2023bbf,Antinucci:2023ezl,Bhardwaj:2023fca}.

Since topological quantum field theories are much more tractable than ordinary quantum field theories, the SymTFT is a powerful tool for probing symmetries and their anomalies. E.g. in low dimensions TQFTs can be classified and, in turn, this becomes a classification of the SymTFT constraining possible realization of symmetries in one dimension lower. However, usual axioms of a TQFT as postulated by Atiyah require it to assign a finite dimensional Hilbert space to every closed manifold, this approach naively only works for finite symmetries. Describing continuous symmetries (and even infinite discrete symmetries like $\mathbb{Q} / \mathbb{Z}$) requires a relaxation of the standard TQFT axioms. Such generalizations of the standard TQFT axioms have been studied in various contexts, however the correct generalization for describing continuous symmetries is still under development (for a general overview of TQFT see \cite{BIRMINGHAM1991129}). Physically such a generalization is highly desirable, and recently some proposal were put forth \cite{Antinucci:2024zjp,Brennan:2024fgj, Apruzzi:2024htg, Arbalestrier:2024oqg, Bonetti:2024cjk}. These theories are important data points for developing an axiomatic generalization and hence deserve further study.

In the specific case of Maxwell theory, the proposed SymTFT is a generalization of $5$ dimensional BF theory where the dynamical fields are not gauge field for a finite group, but rather they are $\mathbb{R}$ connections. The scope of this work is to investigate further these theories. We aim at providing new insight on their structure, using parallels with the symmetries of quantum mechanical system to study the automorphisms of said SymTFT, i.e. electromagnetic dualities of Maxwell theory. This analysis will motivate a new proposal for continuous non-invertible symmetries, that restore classical symmetries broken at the quantum level. This procedure, however, will be shown to be consistent only at the price of introducing particular projectors that trivialize Wilson and 't Hooft lines of the dynamical theory. This suggests that classical continuous symmetries can be recovered in the quantum regime as non-invertible ones, at the price of trivializing all non-local operators of the theory.

This work is organized as follows: in \cref{sec:brief-review-symtft} we introduce the proposed SymTFT for the Maxwell theory and the group $SL_2(\mathbb{R})$ of its automorphisms. We derive its action on the SymTFT and its boundary conditions by utilizing the Lie algebra of the operators in the theory. In \cref{sec:noninvsym} we use this action to construct non-invertible symmetries corresponding to the $U(1)$ subgroup of $SL_2(\mathbb{R})$ that stabilizes the gauge coupling $\tau$ and discuss the action of these symmetries on line operators of the Maxwell theory. Consistency of the fusion rules for this symmetry will force the introduction of special kind of condensates that acts on the theory by projecting out all charged line operators. We end with a summary of results and outlook for the future in \cref{sec:conclusions-outlook}.

{\it Note: while this paper was in preparation, \cite{Arbalestrier:2024oqg} appeared on the ArXiv. The two papers present some complementary descriptions of similar phenomena in the context of non-invertible symmetries. In \cite{Arbalestrier:2024oqg}, non-invertible symmetries are studied in the context of massless QED, while in this paper we tackle Maxwell theory. Both approaches are complementary and lead to similar results.}

% In this paper we discuss how the recent developments on the symmetry tft for continuous symetries \cite{Antinucci:2024zjp} \cite{Brennan:2024fgj} agree with the recent discoveries by \cite{Niro:2022ctq}\cite{Sela:2024okz} about the action of $SL_2(\mathbb{Q})$ on Maxwell theory via non-invertible defects. In particular, we discuss these new non-invertible $SL_2(\mathbb{Q})$ defects in the symmetry tft perspective, finding out a whole new $SL_2(\mathbb{R})$ action on the theory. This implies that at any value of the coupling constant in Maxwell theory we have a new $U(1)$ non-invertible symmetry.

\section{Brief review of SymTFT for Maxwell theory}
\label{sec:brief-review-symtft}

For convenience, in this section we reproduce the discussion of \cite{Antinucci:2024zjp} and \cite{Brennan:2024fgj} about the SymTFT of Maxwell theory. This will give us the chance to fix some notation and discuss some aspects of the SymTFT for continuous symmetries.

As claimed in \cite{Antinucci:2024zjp,Brennan:2024fgj}, the SymTFT for Maxwell theory is given by a BF-like theory
\begin{equation}
  \label{eq:4}
  S_{5d} = \frac{1}{2 \pi}\int_{M_5} a \wedge \dd  b \, ,
\end{equation}
where $a$ and $b$ are two-form $\mathbb{R}$ connections, i.e. differential two forms.

Similarly to ordinary BF-theory, one can study the solutions to the equations of motion, i.e. $\dd a = \dd b = 0$, which constrain the $a$ and $b$ to be closed forms. Thus one can always choose representatives $a, b \in H^2(M_5, \mathbb{R})$. The space of classical solutions is thus $H^2(M_5, \mathbb{R}) \oplus H^2(M_5, \mathbb{R})$.

We can now define the observables of the theory
\begin{equation}
    W(M_2)_\alpha = e^{i \alpha \oint_{M_2}a} \ , \ V(N_2)_\beta = e^{i \beta \oint_{N_2}b},
\end{equation}
with $M_2, N_2\in H_2(M_5,\mathbb{R})$. It is easy to see that these operators are well defined and gauge invariant for any choice of $\alpha$, $\beta \in \mathbb{R}$.

These operators also satisfy the commutation relation
\begin{equation}\label{eq:heisengroup}
    W(M_2)_\alpha V(N_2)_\beta = e^{2\pi i \alpha \beta \langle M_2, N_2 \rangle} V(N_2)_\beta W(M_2)_\alpha,
\end{equation}
where $\langle , \rangle$ is the linking form on $M_5$. Because of this relation, we can interpret the $W$ and $V$ operators as generators of a Heisenberg group. Since this is now a continuous Lie group, with parameters $\alpha$ and $\beta$, we can explicitly study its Lie algebra. Taking derivatives of the group elements, we obtain that its Heisenberg algebra is generated by $\oint_{M_2} a$ and $\oint_{N_2} b$. In particular, we can define the following functions
\begin{equation}
    x_{M_2} (a,b) = \oint_{M_2} a \ , \ p_{N_2} (a,b) = \oint_{N_2} b,
\end{equation}
on the space of classical solutions of the SymTFT. They satisfy the commutation relation
\begin{equation}\label{eq:commutation}
    [x_{M_2} , p_{N_2}] =  2 \pi i \langle M_2 , N_2 \rangle \, .
\end{equation} 

These functions, upon quantization, are the analog of “position" and “momentum" operators in quantum mechanics. In the following, we will use this observation to construct the automorphisms group of the SymTFT.

\subsection{Boundary conditions and automorphisms}

As discussed in \cite{Antinucci:2024zjp}, there are two sets of inequivalent boundary conditions, i.e. two choices of sets of mutually commuting operators in the SymTFT. This can be seen easily from the commutation relations that follows from the E.O.M.
\begin{align}
    & W(M_2)_\alpha V(N_2)_\beta = e^{2\pi i \alpha \beta \langle M_2, N_2 \rangle} V(N_2)_\beta W(M_2)_\alpha \, , \nonumber \\
    & W(M_2)_\alpha W(N_2)_\beta = W(N_2)_\beta W(M_2)_\alpha \, , \quad V(M_2)_\alpha V(N_2)_\beta = V(N_2)_\beta V(M_2)_\alpha \, .
\end{align}
We now easily see that two operators commute either if they are of the same kind, i.e. both are, say, $W_\alpha$ and $W_\beta$ over any support, or if $\alpha \beta \in \mathbb{Z}$, since $\langle M_2, N_2 \rangle$ is an integer.\footnote{More generally, if one considers operators of the form $X_{(\alpha,\beta)}=e^{i \alpha x_{M_2} + i \beta p_{M_2}}$ and their commutation relations, one will reach the same conclusion.} Before considering the most general case, let us analyse the ones with the simplest interpretation in terms of the bulk theory
\begin{itemize}
    \item Algebra generated by $W_\alpha$ (analogous for $X_{(\alpha,\beta)}$):

    This condition consists in imposing Dirichlet boundary conditions (DBCs) for $W_\alpha$ and Neumann (NBCs) for $V_\beta$. Upon squeezing the SymTFT, this boundary condition realizes a 1-form $\mathbb{R}$ symmetry implemented by the $V_\beta$ operators i.e. the operators which cannot end on the topological boundary. The $W_\alpha$ operators, instead, can end on the topological boundary and on line operator of the dynamical theory.

    We will call this theory, the $\mathbb{R}$-Maxwell theory.

    \item Algebra generated by $W_m$ and $V_n$:

    The operators allowed to end on the topological boundary are now $W_m$ and $V_n$ with $m,n \in \mathbb{Z}$, which will also end on the Wilson and 't Hooft lines of the dynamical theory. The theory then admits two $U(1)$ 1-form symmetries with integrally quantized charges. These symmetries are implemented by $W_{\hat \alpha}$ and $V_{\hat \beta}$. 

    This theory will be dubbed from now on $U(1)$-Maxwell theory.

\end{itemize}
Any other boundary condition is related to one of these by the $SL_2(\mathbb{R})$ action as we will explain shortly.

Having discussed the interpretation of the gauge invariant operators of the SymTFT in terms of the symmetries they implement on the dynamical theory, we now turn to the discussion of the automorphisms, or 0-form symmetries, of the SymTFT, explaining their action on the boundary conditions and the consequences for the dynamical theories. Indeed, ordinary symmetries of the SymTFT can be interpreted as -1-form symmetries of the dynamical one, since their action changes the global variant and in general has non-trivial effects on the parameters defining the theory.

To describe the automorphisms of the SymTFT, we can use the fact that the operators satisfy commutation relations \cref{eq:heisengroup} describing a Heisenberg group. The same group structure is found in the study of the free quantum particle in 1d, where Stone-von Neumann theorem states that all irreducible representation of this Heisenberg group are unique up the action of $SL_2(\mathbb{R})$. This action furnishes a unitary and faithful representation of $SL_{2}(\mathbb{R})$. Therefore we can identify $SL_2(\mathbb{R})$ as the group of automorphisms of the Heisenberg algebra.

To construct these automorphisms we use the fact that the $sl_2(\mathbb{R})$ algebra can be expressed in terms of the generator of the Heisenberg algebra,\cite{Carter_MacDonald_Segal_1995,Naber+2021}:
\begin{equation}
    \frac{1}{2} p^2 \to
    \begin{pmatrix}
    0 & 1 \\
    0 & 0 
    \end{pmatrix}  \ , \ 
    \frac{1}{2} x^2 \to
    \begin{pmatrix}
    0 & 0 \\
    1 & 0 
    \end{pmatrix}  \  , \ 
    \frac{1}{2} \{x , p \} \to
    \begin{pmatrix}
    1 & 0 \\
    0 & -1 
    \end{pmatrix} \, .
\end{equation}
By a straightforward generalization, the generators of the group of 0-form symmetries of the SymTFT, and the associated $SL_2(\mathbb{R})$ elements, are
\begin{align}\label{eq:automheis}\begin{gathered}
T_a = \exp\left(i \frac{a}{4 \pi} \sum_{i,j = 1}^{b_2} p_{M_2^i} p_{M_2^j} Q_{ij} \right) \, \to
\begin{pmatrix}
    1 & a \\
    0 & 1 
\end{pmatrix} \, , \\ 
U_b = \exp\left(i \frac{b}{4 \pi} \sum_{i,j = 1}^{b_2} x_{M_2^i} x_{M_2^j} Q_{ij} \right) \, \to
\begin{pmatrix}
    1 & 0 \\
    b & 1 
\end{pmatrix} \, , \\ 
G_c = \exp\left(-i \frac{c}{4 \pi} \sum_{i,j = 1}^{b_2} \{ x_{M_2^i} , p_{M_2^j} \} Q_{ij} \right) \, \to
\begin{pmatrix}
    1/c & 0 \\
    0 & c 
\end{pmatrix} \, ,
\end{gathered}
\end{align}
with $a,b,c \in \mathbb{R}$. The $M_2^i$ are a basis of $H_2(M_4, \mathbb{R}) \cong \mathbb{R}^{b_2} $ and $Q_{ij}$ is the intersection pairing of the compact 4-manifold on which the operator has support. This definition is manifestly invariant under any change of basis that preserves the intersection form of $M_{4}$ and thus invariant under diffeomorphisms of $M_{4}$. The adjoint action of these generators on the basis of the algebra, $(x := x_{M_2}, p:= p_{N_2})$, is given by matrix multiplication, for example
\begin{align}
    T_{-a} \begin{pmatrix}
    x  \\
    p 
\end{pmatrix} T_{a} = \begin{pmatrix}
    1 & a \\
    0 & 1 
\end{pmatrix} \cdot \begin{pmatrix}
    x \\
    p 
\end{pmatrix} = \begin{pmatrix}
    x + a p \\
    p 
\end{pmatrix} \, ,
\end{align}
and the action on $W_\alpha$ and $V_\beta$ follows similarly. 

We can now discuss how these symmetries of the SymTFT act on boundary conditions. We start with the $\mathbb{R}$-Maxwell theory where we choose DBC for the $W(M)_\alpha = e^{i \alpha x}$ operator. In this case, it easy to see that the action of $T_a$ and $G_c$ leave the condition unchanged. Indeed, we have that
\begin{align}
T_a (1 , 0) = (1 , 0) \, , \quad G_c (1 , 0)= (1/c , 0)  \, .
\end{align}
However, the action of $U_b$ changes the BC. Since
\begin{align}
U_b (1 , 0) = (1 , b) \, ,
\end{align}
this action results in a DBC for $X_{(\alpha, b \beta)} = e^{ i \alpha x + i b \beta p}$. This still leads to an $\mathbb{R}$-Maxwell theory. In general, starting from an $\mathbb{R}$-Maxwell theory and acting with an element of $SL_2(\mathbb{R})$, one ends up with another $\mathbb{R}$-Maxwell. $\mathbb{R}$-Maxwell boundary conditions form an orbit under the automorphisms of the SymTFT.

The case of $U(1)$-Maxwell is more involved and necessitates more care. First of all we can notice that if $a$ and $b$ are integers, the action of $T_a$ and $U_b$ not only does not change the kind boundary conditions, it also preserves the quantization of charges which remain integral. In fact, it turns out that $T_1$ and $U_1$ are the generators of the discrete subgroup $SL_2(\mathbb{Z})$, which is the duality group of $U(1)$-Maxwell theory. Non-integer values of $a$ and $b$ can be obtained by composing $T_1$ and $U_1$ with $G_c$\footnote{Indeed $T_a= G_{1/\sqrt{\abs{a}}} T_{\text{sign}(a)} G_{\sqrt{\abs{a}}} $ and $U_b= G_{\sqrt{\abs{b}}} U_{\text{sign}(b)} G_{1/\sqrt{\abs{b}}} $.}, thus to complete our analysis we only have to discuss the effect of $G_c$ on the B.C. 

From \cref{eq:automheis} we have that
\begin{align}
    mx \to m \frac{x}{c} , \nonumber \\
    np \to n c p \, .
\end{align}
This action preserves the commutation relation \cref{eq:heisengroup}, but now the coefficients of the generators of the Heisenberg group are, in general, no longer integers. In \cite{Antinucci:2024zjp} it was argued that this fact can be interpreted as a rescaling of the coupling constant $\tau \to \tau/c^2$. We will show that this is indeed the case, since the action of $G_c$, with $c \in \mathbb{Q}$, can be interpreted as a discrete gauging of a non-anomalous subgroup of the global $U(1)^2$ 1-form symmetry. However, when $c \in \mathbb{R} \backslash \mathbb{Q}$, one cannot interpret this as a ``discrete" irrational gauging, but, as we will show, one can make sense of it, matching its action on the coupling with an infinite sequence of gauging operations in the bulk theory.

\subsection{Effect of the Discrete Gauging}
\label{sec:effect-discr-gaug}
In order to make sense of the action of $G_c$ on the $U(1)$-Maxwell's boundary condition, we first make a small detour and we discuss the effect of gauging a non-anomalous $\mathbb{Z}_{N_e} \times \mathbb{Z}_{N_m}$\footnote{The ``no-anomaly" condition is discussed in \cite{Cordova:2023ent} and boils down to requiring $\gcd(N_e,N_m)=1$.} subgroup of the $U(1)^2$ global 1-form symmetry. The effect of this discrete gauging will be a rescaling of the coupling constant, $\tau= \frac{4\pi i}{e^2} + \frac{\theta}{2\pi}$, of the theory, \cite{Cordova:2023ent, Niro:2022ctq}. To this end, we will mainly follow \cite{Witten:1995gf} and appendix A of \cite{Hayashi:2022fkw}. 

Let us consider the partition function of $U(1)$-Maxwell theory
\begin{equation}
    Z[\tau] = \int \mathcal{D}A e^{\frac{i}{2e^2}\int_\mathcal{M} F \wedge \star F + i\frac{\theta}{8\pi^2}\int_\mathcal{M} F\wedge F} \, .
\end{equation}

We start by regularizing the theory, putting it on a lattice so that differential k-forms belong to a finite-dimensional vector space of dimension $B_k$. Furthermore, we denote by $b_k$ the k-th Betti number of the spacetime manifold, which we assume to be torsion-free. After the regularization, the continuous limit can be safely restored, as discussed in \cite{Witten:1995gf}, at the price of adding a UV cut-off 
\begin{equation}
    Z[\tau] = (\mathrm{Im} \tau )^{\frac{1}{2}(B_1 - B_0)} \int \mathcal{D}A e^{\frac{i}{2e^2}\int_\mathcal{M} F \wedge \star F + i\frac{\theta}{8\pi^2}\int_\mathcal{M} F\wedge F} \, .
\end{equation}

Let us now perform a $\mathbb{Z}_{N_e}\times\mathbb{Z}_{N_m}$ gauging by coupling the $U(1)_e\times U(1)_m$ 1-form symmetries to $\mathbb{Z}_{N_{e,m}}$ background fields $B_{e,m}$. Upon summing over these backgrounds, we have the following
\begin{equation}
\begin{split}
        \frac{|H^0(\mathcal{M}, \mathbb{Z}_{N_e})|}{|H^1(\mathcal{M}, \mathbb{Z}_{N_e})|}\frac{|H^0(\mathcal{M},\mathbb{Z}_{N_m})|}{|H^1(\mathcal{M},\mathbb{Z}_{N_m})|}\sum_{B_m\in H^2(\mathcal{M},\mathbb{Z}_{N_m})} \sum_{B_e\in H^2(\mathcal{M}, \mathbb{Z}_{N_e})} (\mathrm{Im} \tau )^{\frac{1}{2}(B_1 - B_0)} \\ \int \mathcal{D}A
        e^{\frac{i}{2e^2}\int_\mathcal{M}( F -B_e )\wedge \star (F-B_e) + i\frac{\theta}{8\pi^2}\int_\mathcal{M} (F-B_e)\wedge (F-B_e) +\frac{i}{2\pi}\int_\mathcal{M}(F-B_e)\wedge B_m} \ .
\end{split}
\end{equation}
First, we perform the sum over magnetic backgrounds, i.e. we evaluate 
\begin{equation}    \frac{|H^0(\mathcal{M},\mathbb{Z}_{N_m})|}{|H^1(\mathcal{M},\mathbb{Z}_{N_m})|}\sum_{B_m\in H^2(\mathcal{M},\mathbb{Z}_{N_m})}e^{\frac{i}{2\pi }\int (F-B_e)\wedge B_m} = \delta\Big(\oint\frac{F-B_e}{2\pi} \in N_m \mathbb{Z} \Big) N_m^{b_0-b_1 +b_2} \ ,
\end{equation}
where we have used the fact that $N_{e,m}^{b_i} = |H^i(\mathcal{M}, \mathbb{Z}_{N_{e,m}})|$.

To see the effects of the Dirac delta above, it is convenient to decompose the field strength as $F = da + m$, where $a$ is a globally defined 1-form, and $m \in H^2(\mathcal{M},\mathbb{Z})$ is an integral class that computes the flux of $F$ through closed surfaces, $\oint F /2\pi \in \mathbb{Z}$. We see that the Dirac delta constrains both the flux of $F$ and $B_e$ to be multiples of $N_m$. In particular we have that $m = N_m \hat{m}$, with $\hat{m} \in H^2(\mathcal{M},\mathbb{Z})$, and $B_e = N_m \hat{B}_e$, where $\hat{B}_e\in H^2(\mathcal{M}, \mathbb{Z}_{N_e})$. Now, in order to have again a properly quantized field strength, we perform the change of variable $a \to N_m a $. According to the split of the field strength, the path integral measure decomposes as $\int\mathcal{D}A = \sum_{m\in H^2(\mathcal{M},\mathbb{Z})}\int\mathcal{D}a$, thus the Jacobian factor is $\mathcal{D}a \to N_m^{B_1 -b_1 -B_0 + b_0}\mathcal{D}a$. As a result, the effect of summing over the magnetic backgrounds rescales the whole action $(F-B_e)\to N_m (\hat{F}-\hat{B}_e)$, where from now on we will drop the hat.\footnote{Notice that, since we assumed $\gcd(N_e,N_m)=1$, the rescaling of $B_e$ by $N_m$ only change the representative $B_e$ of $H^2(\mathcal{M},\mathbb{Z}_{N_e})$. Since we are summing over all $B_e$, this change has no effects.}

The path integral now reads
\begin{equation}
\begin{split}
        N_e^{b_0 - b_1} N_m^\chi   \sum_{B_e\in H^2(\mathcal{M}, \mathbb{Z}_{N_e})}  (\mathrm{Im} (\tau N_m^2) )^{\frac{1}{2}(B_1 - B_0)} \int \mathcal{D}A \\ 
        e^{i \frac{N_m^2}{2e^2}\int_\mathcal{M}( F -B_e )\wedge \star (F-B_e) + i\frac{\theta N_m^2}{8\pi^2}\int_\mathcal{M} (F-B_e)\wedge (F-B_e) } \ ,
\end{split}
\end{equation}
where $2b_0 -2 b_1 + b_2 = \chi$ is the Euler characteristic of $\mathcal{M}$.

We now perform the sum over electric backgrounds. To do so, we proceed as before by decomposing $F = da + m$. The sum over topological sectors depends only on the difference $m-B_e$ which defines an element of $H^2(\mathcal{M},\mathbb{Z}/N_e)$, i.e. $\hat{m}/N_e = m - B_e$, with $\hat{m}$ an integer class. This fact can be interpreted as a fractionalization of fluxes of the theory, i.e. $F$ has no longer integer fluxes in unit of $\tau$ after gauging, but fractional ones, i.e. $\oint F /2\pi \in \frac{\mathbb{Z}}{N_e}$. 

The above sum over the electric backgrounds can be rewritten as
\begin{equation}
\begin{split}
   N_e^{b_0 - b_1} N_m^\chi   \sum_{B_e\in H^2(\mathcal{M}, \mathbb{Z}_{N_e})}  (\mathrm{Im} (\tau N_m^2) )^{\frac{1}{2}(B_1 - B_0)} \int \mathcal{D}A \\ 
        e^{i \frac{N_m^2}{2e^2}\int_\mathcal{M}( da +\frac{\hat{m}}{N_e} )\wedge \star (da +\frac{\hat{m}}{N_e}) + i\frac{\theta N_m^2}{8\pi^2}\int_\mathcal{M} (da +\frac{\hat{m}}{N_e})\wedge (da +\frac{\hat{m}}{N_e}) } \ .
\end{split}
\end{equation}
We can now perform safely the change of variables $a \to a / N_e$, since this is a global 1-form and it doesn't effect the period of the fluxes. The measure get rescaled by the factor $\mathcal{D}a \to N_e^{-(B_1 -b_1 -B_0 + b_0)}\mathcal{D}a$, obtaining 
\begin{equation}
    Z[\tau]_\mathrm{gaugued} = N_m^{\chi} (\mathrm{Im} (N_m^2\tau/N_e^2) )^{\frac{1}{2}(B_1 - B_0)}  \int \mathcal{D}A e^{i \frac{ N_m^2}{2e^2 N_e^2}\int_\mathcal{M} F \wedge \star F + i\frac{\theta N_m^2 }{8\pi^2N_e^2}\int_\mathcal{M} F\wedge F} \ .
\end{equation}
Putting everything together, we have that the effect of gauging a non-anomalous subgroup $\mathbb{Z}_{N_e}\times\mathbb{Z}_{N_m} $, is, up to overall constants, the rescaling of the coupling constant of the theory
\begin{equation}
    Z[\tau]/(\mathbb{Z}_{N_e} \times \mathbb{Z}_{N_m}) = N_m^\chi Z\Big[\frac{N_m^2}{N_e^2}\tau\Big]
\end{equation}

In $U(1)$-Maxwell, symmetry operators are given by $W_{\hat{\alpha}}$  and $V_{\hat{\beta}}$, with $\hat{\alpha}$ and $\hat{\beta}$ in $\mathbb{R}/\mathbb{Z}$. The action of $G_c$ on these operators changes the quantization of their periods, in particular, $\hat{\alpha} \to \hat{\alpha}/c$ and $\hat{\beta} \to  c \hat{\beta}$. From the field theory point of view, this means that the charges of Wilson and 't Hooft loop, respectively, are now measured to be in $\mathbb{Z}/c$ and $c \mathbb{Z}$ respectively. This means that, after the action of $G_c$, the theory has improperly quantized fluxes, in the same way it does after a gauging procedure. Thus, we can interpret the $G_c$ action on boundary conditions as a discrete gauging, for $c=N_m/N_e$, and ultimately, as proposed in \cite{Antinucci:2024zjp}, a rescaling of the coupling constant of the dynamical theory.

\subsection{Action of \texorpdfstring{$SL_2(\mathbb{Q})$}{} on \texorpdfstring{$\tau$}{} and the restoration of \texorpdfstring{$SL_2(\mathbb{R}$)}{}}\label{sec:restorsl2}

From the point of view of the 4d theory Maxwell theory, the subgroup $SL_2(\mathbb{Q})$ of the automorphism of the 5d BF-theory becomes the duality group of the theory. Indeed, when squeezing the 5d SymTFT between the two boundaries in the presence of such defects, the topological boundary conditions are changed and consequently the dynamical Maxwell theory is mapped to another global variant. In particular, if the defects of \cref{eq:automheis} have a topological boundary\footnote{This is achieved by considering \ref{eq:automheis} with $M^i_2 \in H_2(M_4,\partial M_4; \mathbb{R})$, i.e. they are cycles in relative homology.}, such a boundary becomes a topological interface between two Maxwell theories at different couplings \cite{Kaidi:2022cpf}, related by a $SL_2(\mathbb{Q})$ transformation, consistently with the analysis in \cite{Niro:2022ctq, Brennan:2024fgj, Antinucci:2024zjp}. 

\begin{figure}
    \centering
    \includegraphics[scale=0.6]{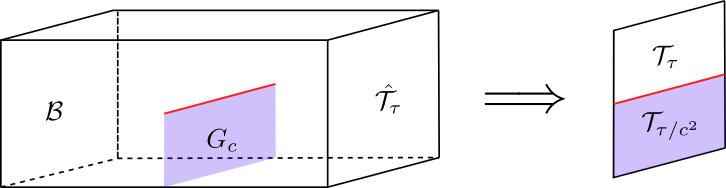}
    \caption{A 0-form symmetry defect of the SymTFT with a boundary becomes an interface between two global variants of the dynamical theory upon shrinking the SymTFT.}\label{fig:the-only-picture}
\end{figure}

Nevertheless, when $c \in \mathbb{R}$, the relation between gauging and $G_c$ action breaks down, since only rational rescaling of the coupling can be achieved via gauging. We thus propose that the full $SL_2(\mathbb{R})$ action can be restored if one allows for countably infinite gauging procedures.

In order to show this, we use the following two facts: first we have that the full $SL_2(\mathbb{R})$ group is generated by $T_1$, $U_1$ and $G_c$, second any irrational number can be expressed as an infinite product of rational ones. This can be achieved as follows: consider a sequence of rational numbers $x_n$ monotonically converging to $c$, then we have that an infinite product approximation of $c$ is given by $c = x_1 \frac{x_2}{x_1} \frac{x_3}{x_2} \dots = z_1 z_2 z_3 \dots$ This can be shown easily using the fact that partial products of the $z_n$ are
\begin{align}
    c_k = a_1 \prod_{i=1}^{k-1} \frac{a_{i+1}}{a_i} = a_k \, ,
\end{align}
thus the sequence of partial products $c_k$ converges to $c$\footnote{In \cref{app:infin-gaug-with} we further expand on how to realize this product using only fractions of prime numbers.}.

\begin{figure}
    \centering
    \includegraphics[scale=0.6]{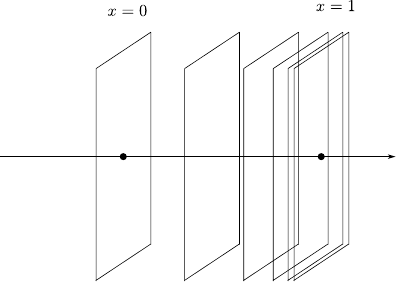}
    \caption{An infinite collection of codimension one operator positioned in the interval $I=[0,1]$ along the normal axis. The n$^{th}$ operator is located at the point $n/(n+1)$, so that none are touching.}\label{fig:the-other-picture}
\end{figure}

It is now easy to see that, since discrete gauging rescales the coupling constant by a rational number, $G_{z_i} (\tau) = \tau/z_i^2$, an infinite number of them can rescale it by any real number. 
% {\color{red} I would remove footnote 5. The quantum dimension of this theory is infinite for sure. However, since this infinite series of operations converge in the SymTFT, we expect to converge also in Maxwell theory. Simply, the TQFT describing this operation does not satisfies Atiyah's axioms.} 
All the operators $G_{z_i} \in SL_{2}(\mathbb{Q})$ are per se well defined.\footnote{As discussed in \cref{app:infin-gaug-with}, they can be constructed avoiding any anomaly.} One can now think of taking a “thick" operators, made of all the $G_{z_i}$ operators in the approximating sequence, put along a closed interval, such that none of them is touching, see \cref{fig:the-other-picture}.
Then, this collection of operators will act on the theory as an $SL_{2}(\mathbb{R})$ duality between the two sides of the thick defect. Finally, we take the (formal) limit in which all the defects coincide and define the resulting operator as the genuine $SL_{2}(\mathbb{R})$ defect.\footnote{Although we cannot prove the convergence of this limit, we view it as natural consequence of the SymTFT described by \eqref{eq:4}. To motivate this, let us consider a closely related theory: a $\mathbb{Q}$ BF-theory. This theory has a $SL(2,\mathbb{Q})$ duality group and it is enough to capture the results of \cite{Niro:2022ctq}. If infinite products do not exist, from the perspective of the dynamical boundary, both $\mathbb{Q}$ and $\mathbb{R}$ BF theories are equivalent, with the difference being the extra $G_c$, with $c$ irrational, operators in the latter. Therefore, either a boundary for the bulk $G_c$ operator doesn't exist, and the two theories cannot be distinguished, or it does. Assuming the latter to be true, and since the passage from $\mathbb{Q}$ to $\mathbb{R}$ consists of adding the infinite products we are considering, we use the above construction to then study the consequences of a bulk $\mathbb{R}$ BF-theory. We would like to thank the anonymous JHEP refree for generous feedback on this point and encouraging us to motivate it better.}
\begin{equation}\label{eq:prodgauge}
G_c(\tau) % \footnote{Let us note that we don't have a proof that this operator has a well defined quantum dimensions. However, being the product of infinitely many one dimensional operators, we are confident that it is well defined. As an example of what we mean, consider infinitely many successive U(1) rotation by angles $2\pi/2^n$, $n \geq 1$. The operator defined as the limiting product of all of these rotation will be the identity, which is indeed a well defined quantum operator.}
=\prod_i G_{z_i}(\tau)=\tau \prod_i 1/z_i^2 = \tau / c^2 \, .
\end{equation}
In a TQFT with a discrete set of operators (e.g. a BF theory with finite gauge group), an infinite product of defect does not always make sense, since that limit might not exist. However, for continuous one this may not be the case. Nevertheless, in the current set up we can match the sequence of gauging on the dynamical boundary with the operator $G_c$ in the bulk of the SymTFT. Since the fields of the the SymTFT are $\mathbb{R}$ connections and the automorphisms group of the theory is the full $SL_2(\mathbb{R})$, thus $G_c$ is a well defined -1-form symmetry of the dynamical theory and can be realized as the limit of infinite gauging as above. When the bulk operator $G_c$ admits a topological boundary, \cref{fig:the-only-picture}, upon shrinking the SymTFT, the operator becomes an interface between two Maxwell theories. Let us note that said boundary, the red line in \cref{fig:the-only-picture}, can be interpreted as higher gauging, in the sense of \cite{Roumpedakis:2022aik}, of $U(1)$ flat connections.\footnote{In this case, we do not have higher anomalies, since the defects we are interested in trivialize all bulk anomaly and, therefore, there is no inflow on the boundary.}\textsuperscript{,}\footnote{We do not have a proof the equivalence of these two approaches. We leave this investigation to future works.}
%\footnote{In the sense that each $\mathcal{O}_{i}$ get closer and closer to $\mathcal{O}$ as elements of $SL_{2}(\mathbb{R})$ and the minimal charge of the operators that isn't projected out becomes higher and higher thus taking us closer and closer to projecting out all lines.} to action of $\mathcal{O}$. 
%These $SL_{2}(\mathbb{Q})$ operators exist in a close cousin of the BF theory we are considering here: namely a BF theory based on gauge group $\mathbb{Q}$. Furthermore, as shown in \cref{app:infin-gaug-with}, one can always choose the infinite sequence of defects implementing the rescaling by an irrational numbers in such a way that the all defects avoid the anomaly discussed in the previous section. Now, a defect implementing an irrational scaling, $\mathcal{O} \in SL_{2}(\mathbb{R})$, can be build a sequence of defects 

Thus, it may seems that we can recover the full action of $SL_2(\mathbb{R})$ on BC in terms of the Maxwell's $SL_2(\mathbb{Z})$ duality plus a possibly infinite number of discrete gauging. However, a more careful analysis shows that this comes at a price.

In the quantum theory, duality defects not in $SL_2(\mathbb{Z})$, enjoy a non-invertible fusion rule, as shown in \cite{Niro:2022ctq, Cordova:2023bja}. As we will show in the next sections, these fusion rules are invertible up to condensates, which in the case of $SL_2(\mathbb{Q})$ dualities/symmetries have a mild effect on the theory, but for $SL_2(\mathbb{R})$ they become projectors that trivialize every line operator in the theory.

\section{The \texorpdfstring{$U(1)$} non-invertible symmetry}\label{sec:noninvsym}

In \cite{Niro:2022ctq, Sela:2024okz}, it was shown that one can construct a (non-)invertible symmetry for a fixed $U(1)$-Maxwell's coupling constant $\tau$. This is because the duality group $SL_2(\mathbb{Q})$ acts on $\tau$ as
\begin{align}
    \tau \to \frac{\tau a + b}{\tau c + d} \, ,
\end{align}
with $a,b,c,d \in \mathbb{Q}$ and $ad-bc=1$. Imposing that $\tau$ remains invariant under the action of $SL(2,\mathbb{Q})$ leads to
\begin{align}\label{eq:stabtau}
    \tau = \frac{a-d + i \text{ sign}(c) \sqrt{4 - (a+d)^2}}{2c} \, ,
\end{align}
with the constraint $-2<(a+d)<2$, such that Im$(\tau)>0$. Thus, the coupling is stabilized by the element of of the duality group associated to the matrix
\begin{align}
    A = \begin{pmatrix}
        a & b \\
        c & d
    \end{pmatrix} \, ,
\end{align}
which will produce a finite, i.e. $\mathbb{Z}_n$, or infinite, i.e.  $\mathbb{Z}$, (non-)invertible symmetry depending on its eigenvalues being, respectively, n$^{th}$-roots of unity or not. This symmetry is nothing but the enhancement of the $SL_2(\mathbb{Q})$ duality group we described in previous sections to a symmetry of the theory, for certain value of $\tau$.

Now, since we expect that via infinite gauging one can recover the full classical $SL_2(\mathbb{R})$ duality group of the theory, it is natural to ask whether the theory can now admit continuous and possibly non-invertible symmetries as well. Indeed, it turns out that every value of $\tau$ is stabilized by a $U(1)$ subgroup of $SL_{2}(\mathbb{R})$. Such a symmetry is present classically, but not at the quantum level since it is inconsistent with the quantization of electric and magnetic fluxes. However, we can nevertheless try to describe it.

For sake of concreteness, suppose that the theory is at a generic $\tau$, one can now perform an $SL_2(\mathbb{R})$ transformation, say $P_{\tau \to i}$, to bring the coupling to $\tau'=P_{\tau \to i}(\tau)=i$. Now, there is an $SO(2) \subset SL_2(\mathbb{R})$ group that stabilize the coupling
\begin{equation}
   \tau' = i \to \frac{
    i a +b}{i c + d} \, ,
\end{equation}
if we choose $a=d=$cos$\theta$ and $c=-b=$sin$\theta$, this acts as $R_\theta(i)=i$. We now act with another $SL_2(\mathbb{R})$ transformation, $P^{-1}_{\tau \to i}, $to bring the coupling back to $\tau$. It is easy to see that any $\tau$ is left invariant by the composition of these transformations, i.e. $P^{-1}_{\tau \to i} ( R_\theta ( P_{\tau \to i}(\tau)))=\tau$. From now on, for sake of simplicity, we fix $\tau = i$.

The defect implementing the rotation, $R_\theta$, in the dynamical theory can be expressed using the operators in \cref{eq:automheis}, with a slight abuse of notation\footnote{$U_n,T_m$ and $G_c$ here indicate the 3 dimensional defects obtained from the 4 dimensional ones defined in \cref{eq:automheis}, with a topological boundary.}, as follows:
\begin{equation}
    R_\theta = U_{\tan{\theta}} T_{-\cos{\theta}\sin{\theta}} G_\frac{1}{\cos{\theta}} \ ,
\end{equation}
for $\cos{\theta}\neq 0$. Furthermore, if $\sin\theta \neq 0$, we can rewrite
\begin{equation}
    U_{\tan{\theta}} = G_{\sqrt{|\tan{\theta}|}} U_{\pm 1} G_\frac{1}{\sqrt{|\tan{\theta}|}} \ ,
\end{equation}
where the plus is needed for $\tan\theta >0$, while the minus is obtained for $\tan\theta <0$, and we can also rewrite (with the same sign conventions as above)
\begin{equation}
    T_{-\cos{\theta}\sin{\theta}} = G_\frac{1}{\sqrt{|\cos{\theta}\sin{\theta}|}} T_{\mp 1} G_{\sqrt{|\cos{\theta}\sin{\theta}|}} \ .
\end{equation}
Since it is required for later discussions, we write the complete formula in the case where $\tan{\theta}>0$
\begin{equation}
   R_\theta = G_{\sqrt{\mathrm{tan}\theta}} U_1 G_{1/\mathrm{sin}\theta} T_{-1}  G_{\sqrt{\mathrm{tan}\theta}} \ ,
\end{equation}
and in the case where $\tan{\theta}<0$
\begin{equation}
     R_\theta = G_{\sqrt{|\mathrm{tan}\theta}|} U_{-1} G_{1/|\sin{\theta}|} T_{1}  G_{\frac{\sqrt{|\sin\theta \cos\theta|}}{\cos\theta}} \ ,
\end{equation}
where $G_{c}$ can be obtained via a series of infinite gauging as explained in the previous section.

One could now conclude, naively, that the theory admits a $U(1)$ global 0-form symmetry. However, as we will discuss in the next sections, the defects implementing the $U(1)$ symmetry enjoy a non-invertible fusion rule. Therefore, before discussing the action of this symmetry on the dynamical theory, we find more instructive to first discuss its fusion rules. From that analysis, it will become clear that this classical symmetry can be restored, at the quantum level, at the price of introducing in the theory some special condensates that act by projecting out all the line operators of the theory.

\subsection{Fusion and continuous condensates}
\label{sec:fusi-cont-cond}

We now address the issue of the fusion rules of these operators, starting by considering fusion of discrete gauging operators. Following \cite{Cordova:2023ent}, we have 
\begin{equation}
    G_{q/n} = \int \mathcal{D}a_1\mathcal{D}a_2 e^{\frac{i}{2\pi}\oint_{\mathcal{M}_3} [ a_1 \wedge (q\dd A_L - n\dd A_R)+ a_2 \wedge (n i\star\dd A_L - q i\star\dd A_R) ] } \ ,
\end{equation}
where $a_1$ and $a_2$ are a $U(1)$ gauge connection and $A_{L/R}$ are the Maxwell's gauge fields respectively on the left and right of the defect. This operator corresponds to gauging a $\mathbb{Z}_q^e\times\mathbb{Z}_n^m$ 1-form electric and magnetic symmetry, which is non-anomalous provided that $\mathrm{gcd}(n,q)=1$.

The fusion of this operator with its “opposite"\footnote{The generic fusion of gauging defects can be found in \cref{app:gauging}.
}, i.e. the operator that gauges a $\mathbb{Z}_n^e\times\mathbb{Z}_q^m$ subgroup, can be computed explicitly. With the convention $\dd A_{I} = F_I$  and $i \star \dd A_{I} = \tilde{F}_I$, the fusion reads 
\begin{align}
\begin{gathered}
    G_{q/n} G_{n/q} =  \int \mathcal{D}a_{1,2}\mathcal{D}b_{1,2} \exp\Big{\{}\frac{i}{2\pi}\oint_{\mathcal{M}_3} \Big{[}a_1 \wedge (q F_L - n F_M) + a_2 \wedge (n \tilde{F}_L - q \tilde{F}_M ) \Big{]} 
    \\
    + \frac{i}{2\pi}\oint_{\mathcal{M}_3} \Big{[}b_1 \wedge (n F_M - q F_R)+ b_2 \wedge (q \tilde{F}_M - n \tilde{F}_R ) \Big{]} \Big{\}} \ ,
\end{gathered}
\end{align}
where we introduced new Maxwell's connections $A_M$ in the region between the two defects. By imposing the equation of motion for $A_M$, we have that
\begin{align}
    n \dd  a_1=n \dd  b_1 \rightarrow a_1 = c + \lambda_n \ , \quad b_1 = c \nonumber
    \\
    q \dd  a_2=q \dd  b_2 \rightarrow a_2 = f + \lambda_q \ , \quad b_2 = f
\end{align}
where $\lambda_l$ is a $\mathbb{Z}_l$ connection. We are thus left with
\begin{align}
\begin{gathered}
    ... = \int \mathcal{D}c\mathcal{D}f\mathcal{D}\lambda \exp  \Big{(} \frac{i}{2\pi}\oint_{\mathcal{M}_3} \Big{(} c \wedge (q F_L - q F_R) +f \wedge (n \tilde{F}_L - n \tilde{F}) + 
    \\
    - q \lambda_n \wedge F_L + n \lambda_q \wedge \tilde{F}_L  \Big{)}\Big{)} \, ,
\end{gathered}
\end{align}
integrating out $c,f$ leads to the identification $A_L = A_R$ up to flat connection, i.e. acting like the identity operator, while the integral over $\lambda_l$ gives the following condensate
\begin{equation}\label{eq:condensate}
     C_{\binom{q}{n}} = \sum_{ \underset{\mathcal{N}_2 \in H_2(\mathcal{M}_3, \mathbb{Z}_q)}{\mathcal{M}_2 \in H_2(\mathcal{M}_3, \mathbb{Z}_n)}
     } \exp \Big{(}{ i q \oint_{\mathcal{M}_2} F + i n \oint_{\mathcal{N}_2} \tilde{F} } \Big{)} \, ,
\end{equation}
where the sum is over $\mathcal{M}_2$ and $\mathcal{N}_2$ cycles, Poincare dual to $\lambda_l$\footnote{We used the fact that $\oint \lambda_l / 2 \pi \in \mathbb{Z}/l$ to fix the factors in the exponential.}.

If one performs multiple gauging operations in sequence and compensates them with their opposite, the end result will be a collection of condensation defects which follow the fusion rule
\begin{equation}
\begin{gathered}
     C_{\binom{q}{n}} C_{\binom{p}{m}} = \sum_{\underset{\mathcal{S}_2 \in \mathbb{Z}_m, \mathcal{T}_2 \in \mathbb{Z}_p}{\mathcal{M}_2 \in \mathbb{Z}_n ,\mathcal{N}_2 \in \mathbb{Z}_q} } \exp \Big{(} i q \oint_{\mathcal{M}_2} F + i n \oint_{\mathcal{N}_2} \tilde{F} + i p \oint_{\mathcal{S}_2} F + i m \oint_{\mathcal{T}_2} \tilde{F} \Big{)} \, \\
     = \sum_{\underset{\mathcal{Y}_2 \in \mathbb{Z}_{\text{lcm}(p,q)}}{\mathcal{X}_2 \in \mathbb{Z}_{\text{lcm}(m,n)}}} \exp \Big{(}i \frac{p m + q n}{\gcd{}(m,n)} \oint_{\mathcal{X}_2} F + i \frac{m p + n q}{\gcd{}(p,q)} \oint_{\mathcal{T}_2} \tilde{F} \Big{)} = \gcd(m,n) \gcd(p,q) C_{\binom{\text{lcm}(p,q)}{\text{lcm}(m,n)}}\, ,
\end{gathered}\footnote{As shown in \cref{app:infin-gaug-with}, one can always choose $q_i$ and $n_i$ such that all the factors are coprime, so all $\gcd$ are 1.}
\end{equation}
where we used the short hand $\mathcal{A}_2 \in \mathbb{Z}_a$ for $\mathcal{A}_2 \in H_2(\mathcal{M}_3, \mathbb{Z}_a)$. Note also that the subscript is not a fraction, no simplification occurs between the two number in parenthesis.

Now, for a rotation $R_\theta \in SL_2(\mathbb{Q})$, we recover the analysis carried out in \cite{Niro:2022ctq}. These rotations correspond to rational angles, thus producing an infinite discrete set of symmetry operators enjoying a non-invertible fusions. In particular, the fusion of one defects and its opposite produce a condensate.

This fact can be easily generalized to the case of rotations by irrational angles if one allows for infinite gauging. As discussed at the end of the previous section, $G_c$ operators, for $c \in \mathbb{R}$, can be realized via a series of infinite gauging and now we know how to deal with the fusion of condensates. We thus define\footnote{There is some abuse of notation here, the operators $G_{c},\overline{G}_{c}$ and $C_{c}$ depend not only on $c$ but also on the sequence $\frac{q_{i}}{n_{i}}$ which converge to $c$.}
\begin{equation}
    G_{c}\overline{G}_{c} = \lim_{n \to \infty}\prod_{i=1}^n G_{q_i/n_i} \prod_{i=n}^1 \overline G_{q_i/n_i} =\lim_{n \to \infty} \prod_{i=1}^n C_{\binom{q_i}{n_i}} \, = C_c \, ,
\end{equation}
where $C_c$ is the “continuous" condensate, corresponding to the condensate of all the $\mathbb{Z}_{q_i}^e \times \mathbb{Z}_{n_i}^m$ symmetry gauged. As before, this can be taken as the formal superposition limit of, otherwise separated, infinitely many condensation defects.

It now easy to compute the fusion rules of two opposite\footnote{The fusion of two generic rotations can be realized using the formulas in \cref{app:gauging} for the fusion of gauging operators and regular matrix multiplication for $T_n$ and $U_n$ operators.} rotations $R_\theta$
\begin{equation}
    R_\theta \overline{R}_\theta = G_{\sqrt{\mathrm{tan}\theta}} U_1 G_{1/\mathrm{sin}\theta} T_{-1}  G_{\sqrt{\mathrm{tan}\theta}}G_{1/\sqrt{\mathrm{tan}\theta}} T_{1} G_{\mathrm{sin}\theta} U_{-1}  G_{1/\sqrt{\mathrm{tan}\theta}} = C_R  ,
\end{equation}
where now $C_R$ is given by
\begin{align}
    C_R = C_{1/\sqrt{\mathrm{tan}\theta}} C_{\mathrm{sin}\theta} C_{\sqrt{\mathrm{tan}\theta}} \, .
\end{align}

For $\theta = \pm \pi/2, \pm \pi$, the above formulas do not hold, but for those values the $U(1)$ rotation is implemented by $ S=T_{-1} U_{1} T_{-1}$ and $C=S^2$ respectively.

\subsection{Action on lines}

Finally, let us address how this continuous symmetry, as well as the continuous condensate, act on Wilson and 't Hooft lines of $U(1)$-Maxwell. At $\tau = i$, we have the following Wilson and 't Hooft loops 
\begin{align}
    W(\gamma)_n = e^{i n \oint_\gamma A} \, , \quad T(\lambda)_m = e^{i m \oint_\lambda \tilde{A}} \, ,
\end{align}
with again d$\tilde{A} = \tilde F = i \star F$ the dual field strength.

Now, the action of $R_\theta$ on $F$ and $\tilde{F}$ is given by
\begin{align}
\begin{cases}\label{eq:rotation}
    F' = \cos\theta F - \sin\theta \tilde{F} \\
    \tilde{F}' = \sin\theta F + \cos\theta \tilde{F}
\end{cases} \, ,
\end{align}
as discussed in \cite{Niro:2022ctq}.\footnote{Let us note that this action on the fields correctly reproduce, as expected, the classical electro-magentic duality one can deduce from the E.O.M. of Maxwell theory in the vacuum.}

From \cref{eq:rotation}, it is is easy to see the action on extended operators, since one can attach a disc to the loops and rewrite them in terms of the field strength 
\begin{align}
    W(\gamma)_n = e^{i n \oint_\Gamma F} \, , \quad T(\lambda)_m = e^{i m \oint_\Lambda \tilde{F}} \, ,
\end{align}
with $\partial \Gamma / \Lambda=\gamma / \lambda$.

We thus have
\begin{align}\label{eq:lineaction}
    R_\theta W(\gamma)_n T(\lambda)_m = W(\gamma)_{n \cos\theta + m \sin\theta} T(\lambda)_{ m \cos\theta - n \sin\theta} R_\theta ,
\end{align}
where if $n \cos\theta + m \sin\theta$ and $ m \sin\theta - n \cos\theta$ are not integers, the loops on the right hand side are improperly quantized and are meaningful only if attached to a disc.

We finish our discussion by addressing the effect of the continuous condensate. A regular condensate acts as a projector of Wilson/'t Hooft operators as follows, \cite{Roumpedakis:2022aik, Niro:2022ctq, Shao:2023gho}, 
\begin{align}
C_{\binom{q}{n}} W^pH^m = 
\begin{cases}
    W^pH^m   & \text{if } p=0 \mod n \text{ and } m=0 \mod q \\
    0  & \text{otherwise}
\end{cases} \, .
\end{align}
This property is a consequence of \cref{eq:condensate}: a condensate passing through a line operator will “measure" it via link of the 1-form operator it is made with the charged line. Thus, 
\begin{align}
C_{\binom{1}{n}} W^p = \sum_{ \mathcal{M}_2 \in H_2(\mathcal{M}_3,\mathbb{Z}_n)} \exp \Big{(}{ i \oint_{\mathcal{M}_2} F } \Big{)}  W^p = W^p \sum_{k=1}^{n} e^{2 \pi i \frac{k}{n} p} \, ,
\end{align}
where the sum now is a discrete delta function, that is zero whenever $p \neq 0 \mod n$. A continuous condensate is made of an infinite sequence of condensates $C_{\binom{q_i}{n_i}}$, where $\prod_i q_i/n_i \in \mathbb{R}$. As commented at the end of \cref{sec:restorsl2}, $c$ can be approximated by a monotonic sequence $x_i$ such that $q_i = x_{i}$, $n_i=x_{i-1}$ and $x_0=1$. This means that for every line of fixed charge $p$, a continuous condensate will eventually have an element $C_{\binom{q_i}{n_i}}$ that projects the line out. Thus, continuous condensates acts as zero projector for every line operator.

Before concluding, let us summarize our discussion: Maxwell theory enjoys, at the quantum level, an $SL_2(\mathbb{Z})$ duality, enhanced to $SL_2(\mathbb{Q})$ if one admits discrete gauging and further enhanced to $SL_2(\mathbb{R})$ if one allows infinite gauging. The theory has a $\mathbb{Z}_2$ symmetry, charge conjugation, and for special value of the coupling, $\tau = i, e^{\frac{2 \pi i}{3}}$, the theory admits an invertible $\mathbb{Z}_{4,6}$ symmetry, respectively. Upon discrete gauging, one can construct infinitely many non-finite discrete symmetries that enjoy a non-invertible fusion rule. Finally, via an infinite number of gauging, one can recover the full classical $U(1)$ symmetry of Maxwell equations i.e. the stabilizer of $\tau$ in $SL_2(\mathbb{R})$, at the price of introducing continuous condensates, that effectively project out all lines operators of the theory, making it effectively a “classical" theory.

\section{Conclusions and outlook}
\label{sec:conclusions-outlook}

In this work we have demonstrated that SymTFT proposed by \cite{Antinucci:2024zjp,Brennan:2024fgj} can be used to better understand topological interfaces in Maxwell theory. This is done by studying the action of $SL_2(\mathbb{R})$ automorphism group of the SymTFT, which now is a continuous group. By exploiting the analogy with free quantum particle systems, we where able to write down explicit expressions for the action of $SL_2(\mathbb{R})$ on both the operators and boundary conditions for the SymTFT. 

Using these automorphisms we have constructed defects for the dynamical theory, by appropriately decorating them with boundary conditions. This leads to topological interfaces between Maxwell theory at different couplings. The salient properties of these interfaces is that they are non-invertible and partially restore, at the quantum level, the classical duality/symmetry of the theory. In particular, since any value of coupling $\tau$ is stabilized by a $U(1)$ subgroup of $SL(2,\mathbb{R})$, these defects give rise to $U(1)$ non-invertible 0-form symmetries in the Maxwell theory. This is a refinement of the work of \cite{Niro:2022ctq} which constructed defects for the rational points of this $U(1)$. Moreover, this is an example of a restoration of a classical symmetry in the quantum regime, which was not broken by anomalies\footnote{We thank Daniel Brennan for insightful discussions on this point.}. 

This is a preliminary work and there are many interesting future directions to explore. Fist of all, a natural question is to investigate the role of continuous non-invertible symmetries in the study of dynamical constraints, such as Ward identities and spontaneous symmetry beaking. Similarly it is important to better understand properties of infinite sequence of discrete gauging of subgroups of $U(1)$ which is required to implement the non-invertible symmetries. Indeed, this procedure requires the presence, as a byproduct, of continuous condensates, which trivialize extended operators of the theory. It would be interesting to better understand how general is this feature, since it seems to pose a strong constraint, i.e. no extended operators, for theories admitting continuous non-invertible symmetries up to condensates, i.e. symmetries obtained via higher-gauging \cite{Roumpedakis:2022aik}. Second, it would be interesting to compare our realization of ``$U(1)$ non-invertible symmetries'' which utilizes the non-trivial topology of the group to those presented in \cite{Arbalestrier:2024oqg} which instead treat it as having discrete topology. Finally, another important application of these ideas will be to apply them to SymTFTs for non-abelian continuous symmetries presented in \cite{Bonetti:2024cjk}.

\section*{Acknowledgments}
We would like to thank Daniel Brennan and Michele Del Zotto for insightful discussions as well as Mahbub Alam for explaining to us the argument proposed in \cref{app:infin-gaug-with}, we also thank Riccardo Argurio, Daniel Brennan, Michele Del Zotto and Raffaele Savelli for careful reading of the draft and much appreciated comments. Finally, we would like to thank the anonymous referee of JHEP for their  thorough reading of the draft, for encouraging us to formulate our arguments better and for finding many typos that were present in the initial draft of this work. The work of AH and DM is supported by funding from the European Research Council (ERC) under the European Union’s Horizon 2020 research and innovation program (grant agreement No.~851931). SM and DM acknowledge support from the Simons Foundation (grant \#888984, Simons Collaboration on Global Categorical Symmetries). DM also acknowledge the support from Swedish Research Council Grant No. 2023-05590.

\appendix

\section{ Classical and Quantum defects}
\label{sec:class-quant-defects}

In \cite{Cordova:2023ent}, the duality defect interpolating between two $U(1)$-Maxwell theories at different couplings is given by
\begin{equation}
    D_{\mathcal{Q}} = \int \mathcal{D}(a,b) e^{\frac{i}{2\pi}\oint_{\mathcal{M}_3} (a, b) \cdot ( (F_L, \tilde{F}_L ) - \mathcal{Q} (F_R, \tilde{F}_R ) )} \ ,
\end{equation}
where $a$ and $b$ are auxiliary fields defined on only on $\mathcal{M}_3$ and $\mathcal{Q}$ is a two by two matrix acting on the vector $(F_R , \tilde{F}_R = i \star F_R )$ from the left.

A crucial role is now played by the fact that $a$ can either be a $\mathbb{R}$ or $U(1)$ connection, leading to different constraints for the coefficients of the matrix $\mathcal{Q}$. 

When $a$ is a $\mathbb{R}$ connection, the coefficients of $\mathcal{Q}$ can take any real value since there are no large gauge transformation that can spoil the gauge invariance of the operator $D_{\mathcal{Q}}$. We will call this a 'classical' defect, since, in general, it does not preserve the quantization condition of the fluxes of the $U(1)$-Maxwell theory.

When $a$ is a $U(1)$ connection, instead, the coefficients of $\mathcal{Q}$ are forced to be integers in order to ensure gauge invariance, thus for $\mathcal{Q} \in SL_2(\mathbb{Z})$ and one recovers the invertible quantum duality defects that preserve the quantization of the Maxwell fluxes. However, this is not the end of the story, as one can further relax the condition on the coefficients of $\mathcal{Q}$, asking that $\mathcal{Q} \in SL_2(\mathbb{Q})$. In this case, the defect $D_{\mathcal{Q}}$ is not gauge invariant, but can be made so by stacking it with a three-dimensional topological theory that absorbs the anomalous gauge transformation, \cite{Kaidi:2021xfk,Choi:2022zal,Choi:2022jqy}. This operation has the effect to make the duality group non-invertible, as discussed in \cite{Niro:2022ctq} and reviewed in \cref{sec:noninvsym}. Furthermore, as discussed in this paper, and in \cite{Arbalestrier:2024oqg} as well, one can further relax the condition on the coefficients and recover the full $SL_2(\mathbb{R})$ duality group, via infinite gauging or via stacking a TFT with a continuous spectrum.

\subsection{Fusion of gauging defects}\label{app:gauging}

We now compute the following fusion
\begin{align}
\begin{gathered}
    G_{q/n} G_{p/m} =  \int \mathcal{D}a_{1,2}\mathcal{D}b_{1,2} \Big{(} \exp\Big{(}\frac{i}{2\pi}\oint_{\mathcal{M}_3} a_1 \wedge (q F_L - n F_M) + a_2 \wedge (n \tilde{F}_L - q \tilde{F}_M )
    \\
    + \frac{i}{2\pi}\oint_{\mathcal{M}_3} b_1 \wedge (p F_M - m F_R)+ b_2 \wedge (m \tilde{F}_M - p \tilde{F}_R) \Big{)}\Big{)} \ ,
\end{gathered}
\end{align}
where we assume $\gcd(p,m)=1$ and $\gcd(q,n)=1$.

We proceed as in \cref{sec:noninvsym}, by integrating out $A_M$. This leads to the following equations for the auxiliary fields $a_i$ and $b_i$
\begin{align}
    n \dd  a_1=p \dd  b_1 \ , \Rightarrow a_1 = \frac{p}{\gcd{(p,n)}} c + \lambda_{\gcd{(p,n)}} \, , \quad b_1 = \frac{n}{\gcd{(p,n)}} c  \, , \nonumber
    \\
    q \dd  a_2=m \dd  b_2 \ , \Rightarrow a_2 = \frac{m}{\gcd{(q,m)}} f + \lambda_{\gcd{(q,m)}} \, , \quad b_2 = \frac{q}{\gcd{(q,m)}} d  \, ,
\end{align}
where $c$ is a $U(1)$ connection and $\lambda_l$ is a $\mathbb{Z}_l$ connections. Plugging these solutions back into the path integral we are left with
\begin{align}
\begin{gathered}
    \int \mathcal{D}c\mathcal{D}f\mathcal{D}\lambda_{\gcd{(q,m)}} \mathcal{D}\lambda_{\gcd{(p,n)}}\exp \Big{(} \frac{i}{2\pi}\oint_{\mathcal{M}_3} \Big{(} c \wedge (\frac{p q}{\gcd{(p,n)}}  F_L- \frac{m n}{\gcd{(q,m)}} F_R) + 
    \\
    + f \wedge (\frac{mn}{\gcd{(n,p)}} (\tilde{F}_L)- \frac{p q }{\gcd{(m,q)}} ( \tilde{F}_R)) + q \lambda_{\gcd{(n,p)}}  \wedge F_L + n \lambda_{\gcd{(m,q)}} \wedge \tilde{F}_L ) \Big{)}\Big{)} \, .
\end{gathered}
\end{align}
The integrals over $\lambda_i$ gives the following condensate
\begin{equation}
     C_{\binom{\gcd{(q,m)}}{\gcd{(p,n)}}} = \sum_{ \underset{\mathcal{N}_2 \in H_2(\mathcal{M}_3, \mathbb{Z}_{\gcd{(q,m)}})}{\mathcal{M}_2 \in H_2(\mathcal{M}_3, \mathbb{Z}_{\gcd{(p,n)}})}}\exp \Big{(} i \oint_{\mathcal{M}_2} q F_L + i \oint_{\mathcal{N}_2} n \tilde{F}_L \Big{)} \, ,
\end{equation}
while the integral over $c,f$ implements the new gauging operator $G_{\frac{pq}{mn}}$, where the fraction should be simplified to its irreducible form. 

Thus the general result for the fusion of gauging defects is
\begin{align}
    G_{q/n} G_{p/m} = G_{\frac{pq}{mn}} \times C_{\binom{\gcd{(q,m)}}{\gcd{(p,n)}}} \, .
\end{align}

\subsection{Infinite gauging without condensation}
\label{app:infin-gaug-with}
To realize the defect $G_{r}$ with $r$ irrational, we need to fuse an infinite sequence of operators $G_{\frac{{p_{i}}}{n_{i}}}$ with $p_{i},n_{i} \in \mathbb{Z}$. If there exists a pair such that $\gcd(p_{i/i+1},n_{i+1/i}) \neq 1$, the fusion product of the sequence will involve a condensate. We now show that one can always avoid this, i.e. for any irrational number $r$, there is always a sequence of rational numbers $\frac{p_{i}}{n_{i}}$\footnote{We always assume that the fraction are in reduced form i.e. $\gcd(p_{i},n_{i})=1$.} such that $\gcd(p_{i},n_{j}) = 1$ for all $i,j$ and
\begin{align}
\label{eq:1}
  \prod_{i} \frac{p_{i}}{n_{i}} = r ~.
\end{align}

To show this, it is more convenient to work with the partial products of the sequence, i.e. the sequence $\tilde{r}_{k}$ defined by,
\begin{align}
\label{eq:2}
  \tilde{r}_{k} = \prod_{i=1}^{k} \frac{p_{i}}{n_{i}} = \frac{\tilde{p}_k}{\tilde{n}_k} ~. 
\end{align}
We can now proceed by iteration, given $\tilde{r}_{k}$ we want to find $\frac{p_{k+1}}{n_{k+1}}$ such that $\tilde{r}_{k+1} \equiv \tilde{r}_{k} \frac{p_{k+1}}{n_{k+1}}$ satisfies $\frac{\tilde{r}_{k} + r}{2} < \tilde{r}_{k+1} < r$, ensuring that the sequence $\tilde{r}_{k}$ converges to $r$. Furthermore $\frac{p_{k+1}}{q_{k+1}}$ must satisfy
\begin{align}
\label{eq:3}
  \gcd(p_{k+1} , \tilde{n}_{k}) = 1 = \gcd(\tilde{p}_{i}, n_{k+1}).
\end{align}

Now, since the set $\mathcal{P}  =\{\frac{p}{n} | p , n \in \mbox{Primes} \}$ is dense in the reals\footnote{This is a non-trivial fact stemming from prime number theorem.}, the set $ I_k = \left(\frac{\tilde{r}_{k}+r}{2 \tilde{r}_{k}}  , \frac{r}{\tilde{r}_{k}}\right) \cap \mathcal{P} \subset \mathbb{R}$ contains infinitely many points of the from $\frac{p}{n}$ with $\gcd(p,n)=1$. Thus, since the number of prime factors of $r_k$ is finite, we can always find two primes $p_{k+1}$ and $n_{k+1}$ satisfying \cref{eq:3} such that $\frac{p_{k+1}}{n_{k+1}}$ is in $I_{k}$. Therefore, it is always possible to construct an approximation of a real number $r$ via a sequence $\tilde{r}_k$ as above, which means that it is always possible to perform a series of infinite gauging whose fusion does not produce condensates\footnote{We thank Mahbub Alam for explaining this proof to us.}.

The sequence described above is not canonical, and in fact there are infinitely many such sequences. As a result the operator $G_{r}$ obtained using this sequence is also not canonical and its fusion rules with other operators depend on the precise sequence chosen, e.g. we can always construct the sequence such that it avoids primes $a$ and $b$. In that case we have $G_{r}G_{\frac{a}{b}} = G_{\frac{ra}{b}}$ and no condensate is produced.

\addcontentsline{toc}{section}{References}

\bibliography{sample1}{} 

\providecommand{\href}[2]{#2}\begingroup\raggedright\begin{thebibliography}{10}

\bibitem{Gaiotto:2014kfa}
D.~Gaiotto, A.~Kapustin, N.~Seiberg and B.~Willett, \emph{{Generalized Global
  Symmetries}}, \href{https://doi.org/10.1007/JHEP02(2015)172}{\emph{JHEP}
  {\bfseries 02} (2015) 172} [\href{https://arxiv.org/abs/1412.5148}{{\ttfamily
  1412.5148}}].

\bibitem{Cordova:2022ruw}
C.~Cordova, T.~T. Dumitrescu, K.~Intriligator and S.-H. Shao, \emph{{Snowmass
  White Paper: Generalized Symmetries in Quantum Field Theory and Beyond}},  in
  \emph{{Snowmass 2021}}, 5, 2022,
  \href{https://arxiv.org/abs/2205.09545}{{\ttfamily 2205.09545}}.

\bibitem{McGreevy:2022oyu}
J.~McGreevy, \emph{{Generalized Symmetries in Condensed Matter}},
  \href{https://doi.org/10.1146/annurev-conmatphys-040721-021029}{\emph{Ann.
  Rev. Condensed Matter Phys.} {\bfseries 14} (2023) 57}
  [\href{https://arxiv.org/abs/2204.03045}{{\ttfamily 2204.03045}}].

\bibitem{Schafer-Nameki:2023jdn}
S.~Schafer-Nameki, \emph{{ICTP lectures on (non-)invertible generalized
  symmetries}},
  \href{https://doi.org/10.1016/j.physrep.2024.01.007}{\emph{Phys. Rept.}
  {\bfseries 1063} (2024) 1}
  [\href{https://arxiv.org/abs/2305.18296}{{\ttfamily 2305.18296}}].

\bibitem{Brennan:2023mmt}
T.~D. Brennan and S.~Hong, \emph{{Introduction to Generalized Global Symmetries
  in QFT and Particle Physics}},
  \href{https://arxiv.org/abs/2306.00912}{{\ttfamily 2306.00912}}.

\bibitem{Bhardwaj:2023kri}
L.~Bhardwaj, L.~E. Bottini, L.~Fraser-Taliente, L.~Gladden, D.~S.~W. Gould,
  A.~Platschorre et~al., \emph{{Lectures on generalized symmetries}},
  \href{https://doi.org/10.1016/j.physrep.2023.11.002}{\emph{Phys. Rept.}
  {\bfseries 1051} (2024) 1}
  [\href{https://arxiv.org/abs/2307.07547}{{\ttfamily 2307.07547}}].

\bibitem{Shao:2023gho}
S.-H. Shao, \emph{{What's Done Cannot Be Undone: TASI Lectures on
  Non-Invertible Symmetry}},
  \href{https://arxiv.org/abs/2308.00747}{{\ttfamily 2308.00747}}.

\bibitem{Carqueville:2023jhb}
N.~Carqueville, M.~Del~Zotto and I.~Runkel, \emph{{Topological defects}},  11,
  2023, \href{https://arxiv.org/abs/2311.02449}{{\ttfamily 2311.02449}}.

\bibitem{Bhardwaj:2017xup}
L.~Bhardwaj and Y.~Tachikawa, \emph{{On finite symmetries and their gauging in
  two dimensions}}, \href{https://doi.org/10.1007/JHEP03(2018)189}{\emph{JHEP}
  {\bfseries 03} (2018) 189}
  [\href{https://arxiv.org/abs/1704.02330}{{\ttfamily 1704.02330}}].

\bibitem{Copetti:2023mcq}
C.~Copetti, M.~Del~Zotto, K.~Ohmori and Y.~Wang, \emph{{Higher Structure of
  Chiral Symmetry}},  \href{https://arxiv.org/abs/2305.18282}{{\ttfamily
  2305.18282}}.

\bibitem{Bhardwaj:2023ayw}
L.~Bhardwaj and S.~Schafer-Nameki, \emph{{Generalized Charges, Part II:
  Non-Invertible Symmetries and the Symmetry TFT}},
  \href{https://arxiv.org/abs/2305.17159}{{\ttfamily 2305.17159}}.

\bibitem{Bartsch:2023wvv}
T.~Bartsch, M.~Bullimore and A.~Grigoletto, \emph{{Representation theory for
  categorical symmetries}},  \href{https://arxiv.org/abs/2305.17165}{{\ttfamily
  2305.17165}}.

\bibitem{Bonetti:2024cvq}
F.~Bonetti, S.~Schafer-Nameki and J.~Wu, \emph{{MTC$[M_3, G]$: 3d Topological
  Order Labeled by Seifert Manifolds}},
  \href{https://arxiv.org/abs/2403.03973}{{\ttfamily 2403.03973}}.

\bibitem{Ji:2019jhk}
W.~Ji and X.-G. Wen, \emph{{Categorical symmetry and noninvertible anomaly in
  symmetry-breaking and topological phase transitions}},
  \href{https://doi.org/10.1103/PhysRevResearch.2.033417}{\emph{Phys. Rev.
  Res.} {\bfseries 2} (2020) 033417}
  [\href{https://arxiv.org/abs/1912.13492}{{\ttfamily 1912.13492}}].

\bibitem{Gaiotto:2020iye}
D.~Gaiotto and J.~Kulp, \emph{{Orbifold groupoids}},
  \href{https://doi.org/10.1007/JHEP02(2021)132}{\emph{JHEP} {\bfseries 02}
  (2021) 132} [\href{https://arxiv.org/abs/2008.05960}{{\ttfamily
  2008.05960}}].

\bibitem{Apruzzi:2021nmk}
F.~Apruzzi, F.~Bonetti, I.~n. Garc\'\i{}a~Etxebarria, S.~S. Hosseini and
  S.~Schafer-Nameki, \emph{{Symmetry TFTs from String Theory}},
  \href{https://doi.org/10.1007/s00220-023-04737-2}{\emph{Commun. Math. Phys.}
  {\bfseries 402} (2023) 895}
  [\href{https://arxiv.org/abs/2112.02092}{{\ttfamily 2112.02092}}].

\bibitem{Baume:2023kkf}
F.~Baume, J.~J. Heckman, M.~H\"ubner, E.~Torres, A.~P. Turner and X.~Yu,
  \emph{{SymTrees and Multi-Sector QFTs}},
  \href{https://doi.org/10.1103/PhysRevD.109.106013}{\emph{Phys. Rev. D}
  {\bfseries 109} (2024) 106013}
  [\href{https://arxiv.org/abs/2310.12980}{{\ttfamily 2310.12980}}].

\bibitem{Freed:2022qnc}
D.~S. Freed, G.~W. Moore and C.~Teleman, \emph{{Topological symmetry in quantum
  field theory}},  \href{https://arxiv.org/abs/2209.07471}{{\ttfamily
  2209.07471}}.

\bibitem{Aharony:2013hda}
O.~Aharony, N.~Seiberg and Y.~Tachikawa, \emph{{Reading between the lines of
  four-dimensional gauge theories}},
  \href{https://doi.org/10.1007/JHEP08(2013)115}{\emph{JHEP} {\bfseries 08}
  (2013) 115} [\href{https://arxiv.org/abs/1305.0318}{{\ttfamily 1305.0318}}].

\bibitem{Kapustin:2014gua}
A.~Kapustin and N.~Seiberg, \emph{{Coupling a QFT to a TQFT and Duality}},
  \href{https://doi.org/10.1007/JHEP04(2014)001}{\emph{JHEP} {\bfseries 04}
  (2014) 001} [\href{https://arxiv.org/abs/1401.0740}{{\ttfamily 1401.0740}}].

\bibitem{Hsin:2018vcg}
P.-S. Hsin, H.~T. Lam and N.~Seiberg, \emph{{Comments on One-Form Global
  Symmetries and Their Gauging in 3d and 4d}},
  \href{https://doi.org/10.21468/SciPostPhys.6.3.039}{\emph{SciPost Phys.}
  {\bfseries 6} (2019) 039} [\href{https://arxiv.org/abs/1812.04716}{{\ttfamily
  1812.04716}}].

\bibitem{Gukov:2020btk}
S.~Gukov, P.-S. Hsin and D.~Pei, \emph{{Generalized global symmetries of $T[M]$
  theories. Part I}},
  \href{https://doi.org/10.1007/JHEP04(2021)232}{\emph{JHEP} {\bfseries 04}
  (2021) 232} [\href{https://arxiv.org/abs/2010.15890}{{\ttfamily
  2010.15890}}].

\bibitem{Kong:2020jne}
L.~Kong, T.~Lan, X.-G. Wen, Z.-H. Zhang and H.~Zheng, \emph{{Classification of
  topological phases with finite internal symmetries in all dimensions}},
  \href{https://doi.org/10.1007/JHEP09(2020)093}{\emph{JHEP} {\bfseries 09}
  (2020) 093} [\href{https://arxiv.org/abs/2003.08898}{{\ttfamily
  2003.08898}}].

\bibitem{Ji:2021esj}
W.~Ji and X.-G. Wen, \emph{{A unified view on symmetry, anomalous symmetry and
  non-invertible gravitational anomaly}},
  \href{https://arxiv.org/abs/2106.02069}{{\ttfamily 2106.02069}}.

\bibitem{DelZotto:2022ras}
M.~Del~Zotto and I.~n. Garc\'\i{}a~Etxebarria, \emph{{Global structures from
  the infrared}}, \href{https://doi.org/10.1007/JHEP11(2023)058}{\emph{JHEP}
  {\bfseries 11} (2023) 058}
  [\href{https://arxiv.org/abs/2204.06495}{{\ttfamily 2204.06495}}].

\bibitem{Kaidi:2022cpf}
J.~Kaidi, K.~Ohmori and Y.~Zheng, \emph{{Symmetry TFTs for Non-invertible
  Defects}}, \href{https://doi.org/10.1007/s00220-023-04859-7}{\emph{Commun.
  Math. Phys.} {\bfseries 404} (2023) 1021}
  [\href{https://arxiv.org/abs/2209.11062}{{\ttfamily 2209.11062}}].

\bibitem{Bashmakov:2022uek}
V.~Bashmakov, M.~Del~Zotto, A.~Hasan and J.~Kaidi, \emph{{Non-invertible
  symmetries of class S theories}},
  \href{https://doi.org/10.1007/JHEP05(2023)225}{\emph{JHEP} {\bfseries 05}
  (2023) 225} [\href{https://arxiv.org/abs/2211.05138}{{\ttfamily
  2211.05138}}].

\bibitem{Kaidi:2023maf}
J.~Kaidi, E.~Nardoni, G.~Zafrir and Y.~Zheng, \emph{{Symmetry TFTs and
  anomalies of non-invertible symmetries}},
  \href{https://doi.org/10.1007/JHEP10(2023)053}{\emph{JHEP} {\bfseries 10}
  (2023) 053} [\href{https://arxiv.org/abs/2301.07112}{{\ttfamily
  2301.07112}}].

\bibitem{Chen:2023qnv}
J.~Chen, W.~Cui, B.~Haghighat and Y.-N. Wang, \emph{{SymTFTs and duality
  defects from 6d SCFTs on 4-manifolds}},
  \href{https://doi.org/10.1007/JHEP11(2023)208}{\emph{JHEP} {\bfseries 11}
  (2023) 208} [\href{https://arxiv.org/abs/2305.09734}{{\ttfamily
  2305.09734}}].

\bibitem{Bashmakov:2023kwo}
V.~Bashmakov, M.~Del~Zotto and A.~Hasan, \emph{{Four-manifolds and Symmetry
  Categories of 2d CFTs}},  \href{https://arxiv.org/abs/2305.10422}{{\ttfamily
  2305.10422}}.

\bibitem{Sun:2023xxv}
Z.~Sun and Y.~Zheng, \emph{{When are Duality Defects Group-Theoretical?}},
  \href{https://arxiv.org/abs/2307.14428}{{\ttfamily 2307.14428}}.

\bibitem{Cordova:2023bja}
C.~Cordova, P.-S. Hsin and C.~Zhang, \emph{{Anomalies of Non-Invertible
  Symmetries in (3+1)d}},  \href{https://arxiv.org/abs/2308.11706}{{\ttfamily
  2308.11706}}.

\bibitem{Bhardwaj:2023idu}
L.~Bhardwaj, L.~E. Bottini, D.~Pajer and S.~Sch\"afer-Nameki, \emph{{Gapped
  Phases with Non-Invertible Symmetries: (1+1)d}},
  \href{https://arxiv.org/abs/2310.03784}{{\ttfamily 2310.03784}}.

\bibitem{Bhardwaj:2023bbf}
L.~Bhardwaj, L.~E. Bottini, D.~Pajer and S.~Schafer-Nameki, \emph{{The Club
  Sandwich: Gapless Phases and Phase Transitions with Non-Invertible
  Symmetries}},  \href{https://arxiv.org/abs/2312.17322}{{\ttfamily
  2312.17322}}.

\bibitem{Antinucci:2023ezl}
A.~Antinucci, F.~Benini, C.~Copetti, G.~Galati and G.~Rizi, \emph{{Anomalies of
  non-invertible self-duality symmetries: fractionalization and gauging}},
  \href{https://arxiv.org/abs/2308.11707}{{\ttfamily 2308.11707}}.

\bibitem{Bhardwaj:2023fca}
L.~Bhardwaj, L.~E. Bottini, D.~Pajer and S.~Schafer-Nameki, \emph{{Categorical
  Landau Paradigm for Gapped Phases}},
  \href{https://arxiv.org/abs/2310.03786}{{\ttfamily 2310.03786}}.

\bibitem{BIRMINGHAM1991129}
D.~Birmingham, M.~Blau, M.~Rakowski and G.~Thompson, \emph{Topological field
  theory},
  \href{https://doi.org/https://doi.org/10.1016/0370-1573(91)90117-5}{\emph{Physics
  Reports} {\bfseries 209} (1991) 129}.

\bibitem{Antinucci:2024zjp}
A.~Antinucci and F.~Benini, \emph{{Anomalies and gauging of U(1) symmetries}},
  \href{https://arxiv.org/abs/2401.10165}{{\ttfamily 2401.10165}}.

\bibitem{Brennan:2024fgj}
T.~D. Brennan and Z.~Sun, \emph{{A SymTFT for Continuous Symmetries}},
  \href{https://arxiv.org/abs/2401.06128}{{\ttfamily 2401.06128}}.

\bibitem{Apruzzi:2024htg}
F.~Apruzzi, F.~Bedogna and N.~Dondi, \emph{{SymTh for non-finite symmetries}},
  \href{https://arxiv.org/abs/2402.14813}{{\ttfamily 2402.14813}}.

\bibitem{Arbalestrier:2024oqg}
A.~Arbalestrier, R.~Argurio and L.~Tizzano, \emph{{The Non-Invertible Axial
  Symmetry in QED Comes Full Circle}},
  \href{https://arxiv.org/abs/2405.06596}{{\ttfamily 2405.06596}}.

\bibitem{Bonetti:2024cjk}
F.~Bonetti, M.~Del~Zotto and R.~Minasian, \emph{{SymTFTs for Continuous
  non-Abelian Symmetries}},  \href{https://arxiv.org/abs/2402.12347}{{\ttfamily
  2402.12347}}.

\bibitem{Carter_MacDonald_Segal_1995}
R.~W. Carter, I.~G. MacDonald and G.~B. Segal, \emph{Lectures on Lie Groups and
  Lie Algebras}, London Mathematical Society Student Texts. Cambridge
  University Press, 1995.

\bibitem{Naber+2021}
G.~L. Naber, \emph{Quantum Mechanics, An Introduction to the Physical
  Background and Mathematical Structure}. De Gruyter, Berlin, Boston, 2021,
  \href{https://doi.org/doi:10.1515/9783110751949}{doi:10.1515/9783110751949}.

\bibitem{Cordova:2023ent}
C.~Cordova and K.~Ohmori, \emph{{Quantum Duality in Electromagnetism and the
  Fine-Structure Constant}},
  \href{https://arxiv.org/abs/2307.12927}{{\ttfamily 2307.12927}}.

\bibitem{Niro:2022ctq}
P.~Niro, K.~Roumpedakis and O.~Sela, \emph{{Exploring non-invertible symmetries
  in free theories}},
  \href{https://doi.org/10.1007/JHEP03(2023)005}{\emph{JHEP} {\bfseries 03}
  (2023) 005} [\href{https://arxiv.org/abs/2209.11166}{{\ttfamily
  2209.11166}}].

\bibitem{Witten:1995gf}
E.~Witten, \emph{{On S duality in Abelian gauge theory}},
  \href{https://doi.org/10.1007/BF01671570}{\emph{Selecta Math.} {\bfseries 1}
  (1995) 383} [\href{https://arxiv.org/abs/hep-th/9505186}{{\ttfamily
  hep-th/9505186}}].

\bibitem{Hayashi:2022fkw}
Y.~Hayashi and Y.~Tanizaki, \emph{{Non-invertible self-duality defects of
  Cardy-Rabinovici model and mixed gravitational anomaly}},
  \href{https://doi.org/10.1007/JHEP08(2022)036}{\emph{JHEP} {\bfseries 08}
  (2022) 036} [\href{https://arxiv.org/abs/2204.07440}{{\ttfamily
  2204.07440}}].

\bibitem{Roumpedakis:2022aik}
K.~Roumpedakis, S.~Seifnashri and S.-H. Shao, \emph{{Higher Gauging and
  Non-invertible Condensation Defects}},
  \href{https://doi.org/10.1007/s00220-023-04706-9}{\emph{Commun. Math. Phys.}
  {\bfseries 401} (2023) 3043}
  [\href{https://arxiv.org/abs/2204.02407}{{\ttfamily 2204.02407}}].

\bibitem{Sela:2024okz}
O.~Sela, \emph{{Emergent non-invertible symmetries in $\mathcal{N}=4$
  Super-Yang-Mills theory}},
  \href{https://arxiv.org/abs/2401.05032}{{\ttfamily 2401.05032}}.

\bibitem{Kaidi:2021xfk}
J.~Kaidi, K.~Ohmori and Y.~Zheng, \emph{{Kramers-Wannier-like Duality Defects
  in (3+1)D Gauge Theories}},
  \href{https://doi.org/10.1103/PhysRevLett.128.111601}{\emph{Phys. Rev. Lett.}
  {\bfseries 128} (2022) 111601}
  [\href{https://arxiv.org/abs/2111.01141}{{\ttfamily 2111.01141}}].

\bibitem{Choi:2022zal}
Y.~Choi, C.~Cordova, P.-S. Hsin, H.~T. Lam and S.-H. Shao,
  \emph{{Non-invertible Condensation, Duality, and Triality Defects in 3+1
  Dimensions}}, \href{https://doi.org/10.1007/s00220-023-04727-4}{\emph{Commun.
  Math. Phys.} {\bfseries 402} (2023) 489}
  [\href{https://arxiv.org/abs/2204.09025}{{\ttfamily 2204.09025}}].

\bibitem{Choi:2022jqy}
Y.~Choi, H.~T. Lam and S.-H. Shao, \emph{{Noninvertible Global Symmetries in
  the Standard Model}},
  \href{https://doi.org/10.1103/PhysRevLett.129.161601}{\emph{Phys. Rev. Lett.}
  {\bfseries 129} (2022) 161601}
  [\href{https://arxiv.org/abs/2205.05086}{{\ttfamily 2205.05086}}].

\end{thebibliography}\endgroup

\end{document}